\documentclass[acmtocl,acmnow]{acmtrans2m}
\usepackage{subfigure}
\usepackage{latexsym}
\usepackage{amsfonts, amssymb, eucal, cmmib57,graphicx}
\usepackage{amsmath, amsxtra}
\usepackage{enumerate}

\def\ent{\vdash}
\newcommand\deff{\mathit{def}}

\def\0{{\mathbf 0}}
\def\1{{\mathbf 1}}
\def\C{{\mathcal C}}
\newcommand{\irule}[2]{\frac{\textstyle\rule[-1.3ex]{0cm}{3ex}#1}%
{\textstyle\rule[-.5ex]{0cm}{3ex}#2}}
\newcommand{\trans}[3]{{#1}\stackrel{#2}{\longrightarrow}{#3}}

\newif\ifremarks
\newcommand{\Commento}[1]{\ifremarks{\bf\footnotesize \$. #1}\fi}

\newtheorem{theorem}{Theorem}[section]

\newtheorem{corollary}[theorem]{Corollary}
\newtheorem{proposition}[theorem]{Proposition}
\newtheorem{lemma}[theorem]{Lemma}
\newdef{definition}[theorem]{Definition}
\newdef{remark}[theorem]{Remark}

\title{Soft Concurrent Constraint Programming}
\author{Stefano Bistarelli \\
 Istituto di Informatica e Telematica,
 C.N.R.,
 Pisa, Italy
 \and
 Ugo Montanari \\
Dipartimento di Informatica,
Universit\`a di Pisa,
  Italy
  \and
Francesca Rossi \\
Dipartimento di Matematica Pura ed Applicata,
Universit\`a di Padova,
Italy.
}

\begin{abstract}
\Commento{
Ampliato e spiegato esempio sezione 5.\\
spostato th. 5 dopo lemma 1, e modificato leggermente\\
lasciata inalterata definizione di cut (vedi perche' su commento apposito)\\
ampliata prova lemma 1.\\
Ugo:manca riferimento a Hoare, smith e egli-milner \cite{powerdomains}.\\
cambiati un po' i theoremi di sezione 6.1\\
aggiunte ultime:\\
vari ritocchi +\\
uniformato $\sqsubseteq$ e $\leq$ che erano usati indiffeentemente\\
uniformato captions (tutte senza punto)\\

}
Soft constraints extend classical constraints to represent
multiple consistency levels, and thus provide a way to express
preferences, fuzziness, and uncertainty. While there are many soft
constraint solving formalisms, even distributed ones, by now there
seems to be no concurrent programming framework where soft
constraints can be handled. In this paper we show how the
classical concurrent constraint (cc) programming framework can
work with soft constraints, and we also propose an extension of cc
languages which can use soft constraints to prune and direct the
search for a solution. We believe that this new programming
paradigm, called soft cc (scc), can be also very useful in many
web-related scenarios. In fact, the language level allows web
agents to express their interaction and negotiation protocols, and
also to post their requests in terms of preferences, and the
underlying soft constraint solver can find an agreement among the
agents even if their requests are incompatible.
\end{abstract}

\category{D.1.3}{Programming Techniques}{Concurrent Programming}[Distributed programming]

\category{D.3.1}{Programming Languages}{Formal Definitions and Theory}[Semantics \and Syntax]

\category{D.3.2}{Programming Languages}{Language Classifications}[Concurrent, distributed, and parallel languages \and Constraint and logic languages]

\category{D.3.3}{Programming Languages}{Language Constructs and Features}[Concurrent programming structures \and Constraints]

\category{F.3.2}{Logics and Meanings of Programs}{Semantics of Programming Languages}[Operational semantics]
\terms{Languages}

\Commento{
Add altri se vuoi tra:
Algorithms;
Design;
Documentation;
Economics;
Experimentation;
Human factors;
Languages;
Legal aspects;
Management;
Measurement;
Performance;
Reliability;
Security;
Standardization;
Theory;
Verification;
}

\keywords{constraints, soft constraints, concurrent constraint programming}

\begin{document}

\begin{bottomstuff}
Research supported
in part by the the the Italian MIUR Projects {\sc cometa} and {\sc napoli} and by
ASI project {\sc ARISCOM}.
\end{bottomstuff}

\maketitle

\section{Introduction}
\label{intro}

The concurrent constraint (cc) paradigm \cite{vijay-book} is a
very interesting computational framework which merges together
constraint solving and concurrency. The main idea is to choose a
{\em constraint system} and use constraints to model communication
and synchronization among concurrent agents.

Until now, constraints in cc were {\em crisp}, in the sense that
they could only be satisfied or violated. Recently, the classical
idea of {\em crisp} constraints has been shown to be too weak to
represent real problems and a big effort has been done toward the
use of soft constraints
\cite{partial-ai,fuzzy1,fuzzy2,prob,schiex-ijcai95,jacm,toplas00,BISTATESI},
which can have more than one level of consistency. Many real-life
situations are, in fact, easily described via constraints able to
state the necessary requirements of the problems. However, usually
such requirements are not hard, and could be more faithfully
represented as preferences, which should preferably be satisfied
but not necessarily. Also, in real life, we are often challenged
with over-constrained problems, which do not have any solution,
and this also leads to the use of preferences or in general of
soft constraints rather than classical constraints.

Generally speaking, a soft constraint is just a classical
constraint plus a way to associate, either to the entire
constraint or to each assignment of its variables, a certain
element, which is usually interpreted as a level of preference or
importance. Such levels are usually ordered, and the order
reflects the idea that some levels are better than others.
Moreover, one has also to say, via suitable combination operators,
how to obtain the level of preference of a global solution from
the preferences in the constraints.

Many formalisms have been developed to describe one or more
classes of soft constraints. For instance consider fuzzy CSPs
\cite{fuzzy1,fuzzy2}, where crisp constraints are extended with a
level of preference represented by a real number between $0$ and
$1$, or probabilistic CSPs \cite{prob}, where the probability to
be in the real problem is assigned to each constraint. Some other
examples are partial \cite{partial-ai} or valued CSPs
\cite{schiex-ijcai95}, where a preference is assigned to each
constraint, in order to satisfy as many constraints as possible,
and thus handle also overconstrained problems.

We think that many network-related problem could be represented
and solved by using soft constraints. Moreover, the possibility to
use a concurrent language on top of a soft constraint system,
could lead to the birth of new protocols with an embedded
constraint satisfaction and optimization framework.

In particular, the constraints could be related to a quantity to
be minimized/maximized but  they could also satisfy policy requirements
given for performance or administrative reasons. This leads to
change the idea of QoS in routing and to speak of {\em
constraint-based} routing
\cite{RFC2702,RFC1102,RajJain00,calisti00}. Constraints are in
fact able to represent in a declarative fashion the needs and the
requirements of agents interacting over the web.

The features of soft constraints could also be useful in
representing routing problems where an imprecise state information
is given \cite{chen98}. Moreover, since QoS is only a specific
application of a more general notion of Service Level Agreement
(SLA), many applications could be enhanced by using such a
framework. As an example consider E-commerce: here we are always
looking for establishing an agreement between a merchant, a client
and possibly a bank. Also, all auction-based transactions need an
agreement protocol. Moreover, also security protocol analysis have
shown to be enhanced by using security levels instead of a simple
notion of secure/insecure level \cite{security-padl01}. All these
considerations advocate for the need of a soft constraint
framework where optimal answers are extracted.

In this paper, we use one of the frameworks able to deal with soft
constraints \cite{ijcai95,jacm}. The framework is based on a
semiring structure that is equipped with the operations needed to
combine the constraints present in the problem and to choose the
best solutions. According to the choice of the semiring, this
framework is able to model all the specific soft constraint
notions mentioned above. We compare the semiring-based framework
with constraint systems {\em ``a la Saraswat''} and then we show
how use it inside the cc framework.
%The use of a
%soft constraint system leads to the use of a cut function able to
%check the soft levels of the constraint store and to transform
%them to a yes/not information of consistency.
%
The next step is the extension of the syntax and operational
semantics of the language to deal with the semiring levels. Here,
the main novelty with respect to cc is that tell and ask agents
are equipped with a preference (or consistency) threshold which is
used to %determine their success, failure, or suspension.
prune the search.

After a short summary of concurrent constraint programming
(\S\ref{sec:cc}) and of semiring-based SCSPs (\S\ref{sec:scsp}),
we show how the concurrent constraint framework can be used to
handle also soft constraints (\S\ref{sec:semf}). Then we integrate
semirings inside the syntax of the language and we change its
semantics to deal with soft levels (\S\ref{sec:scc}). Some notions
of observables able to deal with a notion of optimization and with
{\em success} (\S\ref{sec:success}), {\em fail} (\S\ref{sec:fail})
and {\em hang} computations (\S\ref{sec:hang}) are then defined.
Some examples (\S\ref{sec:example}) and an
application scenario (\S\ref{sec:appl}) conclude our presentation showing
the expressivity of the new language. Finally, conclusions
(\S\ref{sec:concl}) are added to point out the main results and possible
directions for future work.

\section{Background}
\label{sec:back}
\subsection{Concurrent Constraint Programming}
\label{sec:cc} The concurrent constraint (cc) programming paradigm
\cite{vijay-book} concerns the behaviour of a set of concurrent
agents with a shared store, which is a conjunction of constraints.
Each computation step possibly adds new constraints to the store.
Thus information is monotonically added to the store until all
agents have evolved. The final store is a refinement of the
initial one and it is the result of the computation. The
concurrent agents do not communicate directly with each other, but only
through the shared store, by either checking if it entails a given
constraint ({\em ask} operation) or adding a new constraint to it
({\em tell} operation).
\subsubsection{Constraint Systems} A constraint is a relation
among a specified set of variables. That is, a constraint gives
some information on the set of possible values that these
variables may assume. Such information is usually not complete
since a constraint may be satisfied by several assignments of
values of the variables (in contrast to the situation that we have
when we consider a valuation, which tells us the only possible
assignment for a variable). Therefore it is natural to describe
constraint systems as systems of {\em partial} information
\cite{vijay-book}.

The basic ingredients of a constraint system (defined following the
information systems idea) are a set $D$ of {\em primitive
constraints} or {\em tokens}, each expressing some partial
information, and an entailment relation $\vdash$ defined on
$\wp(D) \times D$ (or its extension defined on $\wp(D)\times
\wp(D)$)\footnote{The extension is s.t. $u \vdash v$ iff $u \vdash
P$ for every $P \in v$.} satisfying:
\begin{itemize}
\item $u \vdash P$ for all $P \in u$ (reflexivity) and
\item if $u \vdash v$ and $v \vdash z$, then $u \vdash z$ (transitivity).
\end{itemize}
We also define $u \approx v$ if $u \vdash v$ and $v \vdash u$.

As an example of entailment relation, consider $D$ as the set of
equations over the integers; then $\vdash$ could include the pair
$\langle\{x = 3, x = y\}, y = 3\rangle$, which means that the
constraint $y = 3$ is entailed by the constraints $x = 3$ and $x =
y$. Given $X \in \wp(D)$, let $\overline{X}$ be the set $X$ closed
under entailment. %Consider also $\overline{{\wp (D)}}$ as the set
%of all subsets of $D$ closed under entailment.
Then, a constraint in an information system $\langle \wp (D),
\vdash \rangle$ is simply an element of $\overline{\wp (D)}$.
% (that is, a set of tokens).

As it is well known, %\cite{boh},
$\langle \overline{\wp (D)},
\subseteq\rangle$ is a complete algebraic lattice, the compactness
of $\vdash$ gives the algebraic structure for
$\overline{\wp (D)}$, with
least element $true = \{P \mid \emptyset \vdash P\}$, greatest
element $D$ (which we will mnemonically denote $false$), glbs
(denoted by $\sqcap$) given by the closure of the intersection and
lubs (denoted by $\sqcup$) given by the closure of the union. The
lub of chains is, however, just the union of the members in the
chain. We use $a, b,c, d$ and $e$ to stand for elements of
$\overline{\wp (D)}$; $c \subseteq d$ means $c \vdash d$.
%
%Usually there is also a set $Con \subseteq \overline{\wp (D)}$ which
%contains all {\em consistent} sets of tokens. Such a set has to
%satisfy the following axioms:
%\begin{itemize}
%\item if $u \subseteq v$ and $v \in Con$ then $u \in Con$ (i.e., a
%  subset of a consistent set of tokens is consistent);
%\item if $P \in D$ then $\{P\} \in Con$ (i.e., every single token is
%  consistent);
%\item if $u \vdash P$ then $u \cup \{P\} \in Con$ (i.e., a set of
%  tokens is consistent with all tokens which it entails).
%\end{itemize}
\subsubsection{The hiding operator: Cylindric Algebras} In order
to treat the hiding operator of the language (see later), a general notion of
existential quantifier for variables in
constraints is introduced, which is formalized in terms
of cylindric algebras. This leads to the concept of {\em cylindric
constraint system} over an infinite set of variables $V$ such that
for each variable $x \in V$, $\exists_x : \overline{\wp (D)}
\rightarrow \overline{\wp (D)}$ is an operation satisfying:
\begin{enumerate}
\item $u \vdash \exists_x u$;
\item $u \vdash v$ implies $(\exists_x u) \vdash (\exists_x v)$;
\item $\exists_x(u \sqcup \exists_x v) \approx (\exists_x u) \sqcup (\exists_x v)$;
\item $\exists_x\exists_y u \approx \exists_y\exists_x u$.
\end{enumerate}
%In the following, we assume given a (senumerabln) set of variables
%$V$ with typical elements $x,y,z, \ldots$.
%
\subsubsection{Procedure calls}
In order to model parameter passing, {\em diagonal elements} are
added to the primitive constraints. We assume that, for $x, y$
ranging in $V$, $\overline{\wp (D)}$ contains a constraint
$d_{xy}$ which satisfies the following axioms:
\begin{enumerate}
\item $d_{xx} = true$,
\item if $z \neq x,y$ then $d_{xy} = \exists_z(d_{xz} \sqcup d_{zy})$,
\item if $x \neq y$ then $d_{xy} \sqcup \exists_x(c \sqcup d_{xy}) \ent c$.
\end{enumerate}
Note that the in the previous definition we assume the cardinality
of the domain for $x$, $y$ and $z$ greater than $1$. Note also
that, if $\ent$ models the equality theory, then the elements
$d_{xy}$ can be thought of as the formulas $x = y$.
\subsubsection{The language}
The syntax of a cc program is show in Table~\ref{tab:cc}: $P$ is
the class of programs, $F$ is the class of sequences of procedure
declarations (or clauses), $A$ is the class of agents, $c$ ranges
over constraints, and $x$ is a tuple of variables. Each procedure
is defined (at most) once, thus nondeterminism is expressed via
the $+$ combinator only. We also assume that, in $p(x) :: A$,
we have $vars(A) \subseteq x$, where $vars(A)$ is the set of all variables
occurring free in agent $A$. In a program $P = F.A$, $A$ is the
initial agent, to be executed in the context of the set of
declarations $F$. This corresponds to the language considered in
\cite{vijay-book}, which allows only guarded nondeterminism.
\begin{small}
\begin{acmtable}{250pt}
%\hrule %\vskip2pt
\begin{align*}
P &::= F.A \\
F &::= p(x)::A \mid F.F \\
A &::= success \mid fail \mid tell(c)
\rightarrow A \mid E \mid A \| A \mid \exists_x A \mid p(x)\\
E &::= ask(c) \rightarrow A \mid E+E
\end{align*}
%\vskip2pt
%\hrule
\caption{cc syntax}
\label{tab:cc}
\end{acmtable}
\end{small}

In order to better understand the extension of the language that
we will introduce later, let us remind here the operational
semantics of the agents.
\begin{itemize}
\item agent ``$success$'' succeeds in one step,
\item agent ``$fail$'' fails in one step,
\item agent ``$ask(c) \rightarrow A$'' checks whether constraint $c$
  is entailed by the current store and then, if so, behaves like agent
  $A$. If $c$ is inconsistent with the current store, it fails, and
  otherwise it suspends, until $c$ is either entailed by the current
  store or is inconsistent with it;
\item agent ``$ask(c_1) \rightarrow A_1 + ask(c_2) \rightarrow A_2$''
  may behave either like $A_1$ or like $A_2$ if both $c_1$ and $c_2$ are
  entailed by the current store, it behaves like $A_i$ if $c_i$ only
  is entailed, it suspends if both $c_1$ and $c_2$ are consistent with
  but not entailed by the current store, and it behaves like
  ``$ask(c_1) \rightarrow A_1 $'' whenever ``$ask(c_2 ) \rightarrow
  A_2$'' fails (and vice versa);
\item agent
  ``$tell(c) \rightarrow A$'' adds constraint $c$ to the current store
  and then, if the resulting store is consistent, behaves like $A$,
  otherwise it fails.
%  in an ``eventual'' interpretation of the tell,
%  this same agent adds $c$ to the store (without any consistency
%  check) and then behaves like $A$ (if the resulting store is
%  inconsistent this will result in an uncontrolled behavior of the
%  system, since from now on all ask operations will succeed);
\item agent $A_1 \| A_2$ behaves like $A_1$ and $A_2$ executing
  in parallel;
\item agent $\exists_x A$ behaves like agent $A$, except that the
variables in $x$ are local to $A$;
\item $p(x)$ is a call of procedure $p$.
\end{itemize}
A formal treatment of the cc semantics can
be found in \cite{vijay-book,BP91}.
Also, a denotational semantics of deterministic cc programs, based on
closure operators, can be found in \cite{vijay-book}.
%In \cite{JSS31} a
%{\em don't know} nondeterministic choice instead of the classical
%{\em don't care} is investigated. A particular attention to
%infinite computation is instead considered in \cite{BPP95}.
A more complete survey on several concurrent paradigms
is given also in \cite{BP94}.

\subsection{Soft Constraints}
\label{sec:scsp}

Several formalization of the concept of {\em soft constraints} are
currently available. In the following, we refer to the one based
on c-semirings \cite{jacm,BISTATESI}, which can be shown to
generalize and express many of the others.

A soft constraint may be seen as a constraint where each
instantiations of its variables has an associated value from a
partially ordered set which can be interpreted as a set of
preference values. Combining constraints will then have to take
into account such additional values, and thus the formalism has
also to provide suitable operations for combination ($\times$) and
comparison ($+$) of tuples of values and constraints. This is why
this formalization is based on the concept of c-semiring, which is
just a set plus two operations.
%
%\begin{definition}[semirings and c-semirings]
%
\subsubsection{C-semirings}
A semiring is a tuple $\langle A,+,\times,\0,\1 \rangle$ such
that:
\begin{enumerate}
\item $A$ is a set and $\0, \1 \in A$;
\item $+$ is commutative, associative and $\0$ is its unit element;
\item $\times$ is associative, distributes over $+$, $\1$  is its unit element and
$\0$ is its absorbing element.
\end{enumerate}

A {\em c-semiring} is a semiring $\langle A,+,\times,\0,\1
\rangle$ such that $+$ is idempotent, $\1$ is its absorbing
element and $\times$ is commutative.
%\end{definition}
Let us consider the relation $\leq_S$ over $A$ such that $a \leq_S
b$ iff $a+b = b$. Then it is possible to prove that (see
\cite{jacm}):
\begin{enumerate}
\item $\leq_S$ is a partial order;
\item $+$ and $\times$ are monotone on $\leq_S$;
\item $\0$ is its minimum and $\1$ its maximum;
\item $\langle A,\leq_S \rangle$ is a complete lattice and,
for all $a, b \in A$, $a+b = lub(a,b)$.
\end{enumerate}

Moreover, if $\times$ is idempotent, then: $+$ distribute over
$\times$; $\langle A,\leq_S \rangle$ is a complete distributive
lattice and $\times$ its glb. Informally, the relation $\leq_S$
gives us a way to compare semiring values and constraints. In
fact, when we have $a \leq_S b$, we will say that {\em b is better
than a}. In the following, when the semiring will be clear from
the context, $a \leq_S b$ will be often indicated by $a \leq b$.
\subsubsection{Soft Constraints and Problems}
Given a semiring $S = \langle A,+,\times,\0,\1 \rangle$, a finite
set $D$ (the domain of the variables) and an ordered set of
variables $V$, a {\em constraint} is a pair $\langle
\deff, con \rangle$ where $con \subseteq V$ and $\deff: D^{|con|}
\rightarrow A$. Therefore, a constraint specifies a set of
variables (the ones in $con$), and assigns to each tuple of values
of these variables an element of the semiring. Consider two
constraints $c_1=\langle def_1, con \rangle$ and $c_2=\langle
def_2, con \rangle$, with $|con|=k$. Then $c_1 \sqsubseteq_S c_2$
if for all k-tuples $t$, $def_1(t) \leq_S def_2(t)$. The relation
$\sqsubseteq_S$ is a partial order.

A {\em soft constraint problem} is a pair $\langle C, con \rangle$
where $con \subseteq V$ and $C$ is a set of constraints: $con$ is
the set of variables of interest for the constraint set $C$, which
however may concern also variables not in $con$. Note that a
classical CSP is a SCSP where the chosen c-semiring is: $S_{CSP} =
\langle \{false, true\},\vee, \wedge, false, true \rangle$. Fuzzy
CSPs \cite{schiex} can instead be modeled in the SCSP framework by
choosing the c-semiring $S_{FCSP} = \langle [0,1], max, min, 0, 1
\rangle$. Many other ``soft'' CSPs (Probabilistic, weighted,
\ldots) can be modeled by using a suitable semiring structure
(for example, $S_{prob} = \langle [0,1], max, \times, 0, 1 \rangle$,
$S_{weight} = \langle \mathcal{R}, min, +, 0, +\infty \rangle$,
\ldots).

\begin{figure}[tp]
\centering
    %\leavevmode
\includegraphics[scale=.65]{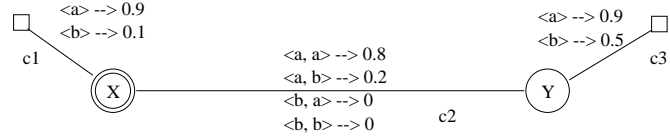}
\caption{A fuzzy CSP}
\label{fig:fuzzy}
\end{figure}

Figure \ref{fig:fuzzy} shows the graph representation of a fuzzy
CSP. Variables and constraints are represented respectively by
nodes and by undirected arcs (unary for $c_1$ and $c_3$ and binary for
$c_2$), and semiring values are written to the right of the
corresponding tuples. The variables of interest (that is the set
$con$) are represented with a double circle. Here we assume that
the domain $D$ of the variables contains only elements $a$ and
$b$.
\subsubsection{Combining and projecting soft constraints}
Given two constraints $c_1 = \langle \deff_1,con_1 \rangle$ and
$c_2 = \langle \deff_2,con_2 \rangle$, their {\em combination}
$c_1 \otimes c_2$ is the constraint $\langle \deff,con \rangle$
defined by $con = con_1 \cup con_2$ and $\deff(t) = \deff_1(t
\downarrow^{con}_{con_1}) \times \deff_2(t
\downarrow^{con}_{con_2})$, where $t\downarrow^X_Y$ denotes the
tuple of values over the variables in $Y$, obtained by projecting
tuple $t$ from $X$ to $Y$. In words, combining two constraints
means building a new constraint involving all the variables of the
original ones, and which associates to each tuple of domain values
for such variables a semiring element which is obtained by
multiplying the elements associated by the original constraints to
the appropriate subtuples.

Given a constraint $c = \langle \deff,con \rangle$ and a subset
$I$ of $V$, the {\em projection} of $c$ over $I$, written
$c\Downarrow_I$ is the constraint $\langle \deff', con' \rangle$
where $con' = con \cap I$ and $\deff'(t') = \sum_{t / t
\downarrow^{con}_{I \cap con} = t'} \deff(t)$.
%\end{definition}
Informally, projecting means eliminating some variables. This is
done by associating to each tuple over the remaining variables a
semiring element which is the sum of the elements associated by
the original constraint to all the extensions of this tuple over
the eliminated variables. In short, combination is performed via
the multiplicative operation of the semiring, and projection via
the additive one.
\subsubsection{Solutions}
The {\em solution} of an SCSP problem
$P = \langle C,con \rangle$ is the constraint $Sol(P)=(\bigotimes
C)\Downarrow_{con}$.
%\end{definition}
That is, we combine all constraints, and then project over the
variables in $con$. In this way we get the constraint over $con$
which is ``induced'' by the entire SCSP.

For example, the solution of the fuzzy CSP of Figure
\ref{fig:fuzzy} associates a semiring element to every domain
value of variable $x$. Such an element is obtained by first
combining all the constraints together. For instance, for the
tuple $\langle a,a \rangle$ (that is, $x=y=a$), we have to compute
the minimum between $0.9$ (which is the value assigned to $x=a$ in
constraint $c_1$), $0.8$ (which is the value assigned to $\langle
x=a,y=a \rangle$ in $c_2$) and $0.9$ (which is the value for $y=a$
in $c_3$). Hence, the resulting value for this tuple is $0.3$. We
can do the same work for tuple $\langle a,b \rangle \rightarrow
0.2$, $\langle b,a \rangle \rightarrow 0$ and $\langle b,b \rangle
\rightarrow 0$. The obtained tuples are then projected over
variable $x$, obtaining the solution $\langle a \rangle
\rightarrow 0.8$ and $\langle b \rangle \rightarrow 0$.

Sometimes it may be useful to find only a semiring value
representing the least upper bound among the values yielded by the
solutions. This is called the {\em best level of consistency} of
an SCSP problem $P$ and it is defined by $blevel(P) = Sol(P)
\Downarrow_{\emptyset}$ (for instance, the fuzzy CSP of Figure
\ref{fig:fuzzy} has best level of consistency $0.8$).
We also say
that: $P$ is $\alpha$-consistent if $blevel(P) = \alpha$; $P$ is
consistent iff there exists $\alpha > \0$ such that $P$ is
$\alpha$-consistent; $P$ is inconsistent if it is not consistent.

\section{Concurrent Constraint Programming over Soft Constraints}
\label{sec:semf} Given a semiring $S = \langle A,+,\times,\0,\1
\rangle$ and an ordered set of variables $V$ over a finite domain
$D$, we will now show how soft constraints over $S$ with a
suitable pair of operators form a semiring, and then, we highlight
the properties needed to map soft constraints over constraint
systems {\em ``a la Saraswat''} (as recalled in
Section~\ref{sec:cc}.

We start by giving the definition of the carrier set of the
semiring.

\begin{definition}[(functional constraints)]
\label{def:carrierset}%Consider a semiring
%$S = \langle A,+,\times,\0,\3 \rangle$, a
%domain of the variables $D$ and an ordered set of variables $V$.
We define %$\C =\{c \mid c:D^{|V|}\rightarrow A\}$
$\C = (V \rightarrow D)\rightarrow A$ as the set of all possible
constraints that can be built starting from $S = \langle
A,+,\times,\0,\1 \rangle$, $D$ and $V$.
%\footnote{This way of representing
%constraints as functions was already given in \cite{cp2050hosw}.}.
\end{definition}
A generic function describing the assignment of domain elements to
variables will be denoted in the following by $\eta : V\rightarrow
D$. Thus a constraint is a function which, given an assignment
$\eta$ of the variables, returns a value of the semiring.

Note that in this {\em functional} formulation, each constraint is
a function and not a pair representing the variable involved and
its definition. Such a function involves all the variables in $V$,
but it depends on the assignment of only a finite subset of them.
We call this subset the {\em support} of the constraint. For
computational reasons we require each support to be finite.

\begin{definition}[(constraint support)]
Consider a constraint $c\in \C$. We define his support as
$supp(c)=\{v\in V \mid \exists \eta,d_1,d_2. c\eta[v:= d_1] \neq
c\eta[v:= d_2]\}$, where
\[\eta[v:= d]v' =
\begin{cases}
d & \text{if $v=v'$},\\
\eta v' & \text{otherwise}.\\
\end{cases}\]
\end{definition}
Note that $c\eta[v:= d_1]$ means $c\eta'$ where $\eta'$ is $\eta$
modified with the association $v:= d_1$ (that is the operator $[\
]$ has precedence over application).

\begin{definition}[(functional mapping)]
Given any soft constraint $\langle def,\{v_1,\ldots,v_n\}\rangle
\in C$, we can define its corresponding function $c\in\C$ s.t.
$c\eta[v_1 := d_1] \ldots [v_n:= d_n]=def(d_1, \ldots, d_n)$.
Clearly $supp(c) \subseteq \{v_1, \ldots, v_n\}$.
\end{definition}

%The definition of support, the following proposition easily holds:
%\begin{theorem}[finite support]

\begin{definition}[(Combination and Sum)]\label{def:combsum}
Given the set $\C$, we can define the combination and sum
functions $\otimes,\oplus: \C\times\C \rightarrow \C$ as follows:
\begin{align*}
(c_1\otimes c_2)\eta &= c_1\eta\times_S c_2\eta \text{\hspace{1cm}
and \hspace{1cm}}
(c_1\oplus c_2)\eta = c_1 \eta+_S c_2\eta.
\end{align*}
\end{definition}
Notice that function $\otimes$ has the same meaning of the already
defined $\otimes$ operator (see Section~\ref{sec:scsp}) while
function $\oplus$ models a sort of disjunction.

By using the $\oplus_S$ operator we can easily extend the partial
order $\leq_S$ over $\C$ by defining $c_1 \sqsubseteq_S c_2 \iff c_1
\oplus_S c_2 = c_2$. In the following, when the semiring will be clear from
the context, we will use $\sqsubseteq$.

We can also define a unary operator that will be useful to
represent the unit elements of the two operations $\oplus$ and
$\otimes$. To do that, we need the definition of constant
functions over a given set of variables.
\begin{definition}[(constant function)]
\label{def:constfunc}
%Consider a semiring $S = \langle
%A,+,\times,\0,\1 \rangle$, a domain of the variables $D$ and an
%ordered set of variables $V$.
We define function $\bar{a}$ as the function that returns the
semiring value $a$ for all assignments $\eta$, that is,
$\bar{a}\eta =a$. We will usually write $\bar{a}$ simply as $a$.
\end{definition}
An example of constants that will be useful later are $\bar{\0}$
and $\bar{\1}$ that represent respectively the constraint
associating $\0$ and $\1$ to all the assignment of domain values.

It is easy to verify that each constant has an empty support. More
generally we can prove the following:
\begin{proposition}
The support of a constraint $c\Downarrow_I$ is always a subset of
$I$(that is $supp(c\Downarrow_I)\subseteq I$).
\end{proposition}
\begin{proof}
By definition of $\Downarrow_I$, for any variable $x \not \in I$
we have $c\Downarrow_I\eta[x=a]=c\Downarrow_I\eta[x=b]$ for any
$a$ and $b$. So, by definition of support $x \not \in
supp(c\Downarrow_I)$.
\end{proof}

%We will
%usually represent $\bar{\0}$ and $\bar{\2}$ as $\0$ and $\1$.

\begin{theorem}[(Higher order semiring)]
The structure $S_C=\langle \C,\oplus,\otimes,\0,\1 \rangle$ where
\begin{itemize}
\item $\C : (V\rightarrow D)\rightarrow A$
is the set of all the possible constraints that can be built
starting from $S$, $D$ and $V$ as defined in
Definition~\ref{def:carrierset},
\item $\otimes$ and $\oplus$ are the functions defined in
Definition~\ref{def:combsum}, and
\item $\0$ and $\1$ are constant functions defined following
Definition~\ref{def:constfunc},
\end{itemize}
is a c-semiring.
\end{theorem}
\begin{proof}
To prove the theorem it is enough to check all the properties with
the fact that the same properties hold for semiring $S$. We give
here only a hint, by showing the commutativity of the $\otimes$
operator:\\
$c_1 \otimes c_2 \eta =$ (by definition of $\otimes$)\\
$c_1\eta \times c_2\eta =$ (by commutativity of $\times$)\\
$c_2\eta \times c_1\eta =$ (by definition of $\otimes$)\\
$c_2 \otimes c_1 \eta$.\\
All the other properties can be proved similarly.
\end{proof}

%\section{Concurrent Soft Constraint Programming}
\label{sec:cscp}

The next step is to look for a notion of token and of entailment
relation. We define as tokens the functional constraints in $\C$
and we introduce a relation $\ent$ that is an entailment relation
when the multiplicative operator of the semiring is idempotent.

\begin{definition}[($\ent$ relation)]
\label{def:ent} Consider the high order semiring carrier set $\C$
and the partial order $\sqsubseteq$. We define the relation $\ent
\subseteq \wp(\C) \times \C$ s.t. for each $C \in \wp(\C)$ and $c
\in \C$, we have $C \ent c \iff \bigotimes C \sqsubseteq c$.
\end{definition}

The next theorem shows that, when the multiplicative operator of
the semiring is idempotent, the $\ent$ relation satisfies all the
properties needed by an entailment.

\begin{theorem}[($\ent$ with idempotent $\times$ is an entailment relation)]
Consider the higher order semiring carrier set $\C$ and the
partial order $\sqsubseteq$. Consider also the relation $\ent$ of
Definition~\ref{def:ent}. Then, if the multiplicative operation of
the semiring is idempotent, $\ent$ is an entailment relation.
\end{theorem}
\begin{proof}
Is enough to check that for any $c \in \C$, and for any $C_1$,
$C_2$ and $C_3$ subsets of $\C$ we have
\begin{enumerate}
\item {\bf $C \ent c$} when $c \in C$: We need to show that
$\bigotimes C \sqsubseteq c$ when $c \in C$. This follows
from the extensivity of $\times$.

\item {\bf if $C_1 \ent C_2$ and $C_2 \ent C_3$ then $C_1 \ent C_3$}:
To prove this we use the extended version of the relation $\ent$
able to deal with subsets of $\C: \wp(\C) \times \wp(\C)$ s.t.
$C_1 \ent C_2 \iff C_1 \ent \bigotimes C_2$. Note that when
$\times$ is idempotent we have that, $\forall c_2 \in C_2, \ C_1
\ent c_2 \iff C_1 \ent \bigotimes C_2$. In this case to prove the
item we have to prove that if $\bigotimes C_1 \sqsubseteq
\bigotimes C_2$ and $\bigotimes C_2 \sqsubseteq \bigotimes C_3$,
then $\bigotimes C_1 \sqsubseteq \bigotimes C_3$. This
comes from the transitivity of $\sqsubseteq$.
\end{enumerate}
\end{proof}

%
%
%\begin{theorem}[Soft Constraint System]
% Let also consider a set of soft constraint
%tokens $\mathcal{C'} \subseteq \C$ and his closure defined as
%$\bar{\mathcal{C'}} = \{c\in\C \mid \bigotimes \mathcal{C'}
%\sqsubseteq c\}$. Then $\C$ and $\sqsubseteq$ describe a
%Constraint System {\em a la Saraswat''}.
%\end{theorem}
%\begin{proof}
%It is enough to observe that the closure of a set functional
%constraints $\bar\wp(\mathcal{C'})$ can in fact be represented by
%an element of $\C$ (represented by $\bigotimes \mathcal{C'}$).
%with this in mind we can use the constraint itself instead of the
%the set and the properties to be satisfied by the $sqsubseteq$
%relations becomes easy to be checked.
%\end{proof}

Note that in this setting the notion of token (constraint) and of
set of tokens (set of constraints) closed under entailment is used
indifferently. In fact, given a set of constraint functions $C_1$,
its closure w.r.t. entailment is a set $\bar{C_1}$ that
contains all the constraints greater than $\bigotimes C_1$. This
set is univocally representable by the constraint function
$\bigotimes C_1$.

%It is important to notice that in a constraint system ``a la
%Saraswat'' the entailment operator is used in a operational
%fashion to incrementally add piece of entailed information to the
%store. This notion can be used in our framework only when the
%times operator is idempotent, because only in this case the
%incrementally added information does not modify the {\em
%semantics} of the store.
%
%This lead to the important result that when the $\times$ operator of
%the semiring is idempotent we obtain a soft constraint system ``a
%la Saraswat''.
%
%\begin{proposition}[Soft Constraint System ``a la Saraswat'']
%Consider the high order semiring carrier set $\C$ defined on top
%of an idempotent semiring operator, and the partial order
%$\sqsubseteq$.  Then $\C$ and $\sqsubseteq$ describe a Constraint
%System {\em ``a la Saraswat''}.
%\end{proposition}
%
%

The definition of the entailment operator $\ent$ on top of the
higher order semiring $S_C=\langle \C,\oplus,\otimes,\0,\1
\rangle$ and of the $\sqsubseteq$ relation leads to the notion of
{\em soft constraint system}. It is also important to notice that
in \cite{vijay-book} it is claimed that a constraint system is a
{\em complete} {\em algebraic} lattice. Here we do not ask for
this, since the algebraic nature of the structure $\C$
strictly depends on the properties of the semiring.\\

%Moreover,  we do not plan (at least not in this work) to
%use the entailment operator to {\em compute} and generate from
%{\em finite} set of tokens new tokens.

%Moreover, since we consider the soft constraint framework
%as defined in \cite{ijcai95,jacm} where the combination operator
%could also be non idempotent, we do not ask for the completeness
%of the lattice. The only properties we ask the soft constraint
%systems to satisfy is to be a lattice.

\subsection{Non-idempotent $\times$}
If the constraint system is defined on top of a non-idempotent
multiplicative operator, we cannot obtain a $\ent$ relation
satisfying all the properties of an entailment. Nevertheless, we
can give a {\em denotational} semantics to the constraint store,
as described in Section~\ref{sec:scc}, using the operations of the
higher order semiring.

To treat the hiding operator of the language, a general notion of
existential quantifier has to be introduced by using notions
similar to those used in cylindric algebras. Note however that
cylindric algebras are first of all boolean algebras. This could
be possible in our framework only when the $\times$ operator is
idempotent.
%This leads to the concept of {\em soft} cylindric constraint
%system.

%\begin{theorem}
%Consider a soft constraint system $SCC = \langle
%\mathcal{C},\sqsubseteq,\otimes,\0_{V},\1_{V}\rangle$. Assume given
%a (denumerable) set of variables $V$ with typical elements $x_i$.
%For each $x_i \in V$ a function $\exists_{x_i} : \mathcal{C}
%\rightarrow \mathcal{C}$ s.t., for each $c \in \mathcal{C}$ we
%have $\exists_{x_i} c(x_1,\ldots,x_i,\ldots,x_{|V|}) =
%\sum_{x_i\in D} c(x_1,\ldots,x_i,\ldots,x_{|V|})$. Then,
%$SCC_{cyl} = \langle \mathcal{C},\sqsubseteq,\otimes,\0_{V},\1_{V},
%V, \exists_x \rangle$ is a cylindric constraint system.
%\end{theorem}
\begin{definition}[(hiding)]
\label{def:hiding}
%Consider a semiring $S = \langle A,+,\times,\0,\1 \rangle$,
Consider a set of variables $V$ with domain $D$ and the
corresponding soft constraint system $\C$. We define for each $x
\in V$ the hiding function
$(\exists_x c)\eta =\sum_{d_i\in D} c\eta[x := d_i]$.
\end{definition}
Notice that $x$ does not belong to the support of $\exists_x c$.

By using the hiding function we can represent the $\Downarrow$
operator defined in Section~\ref{sec:scsp}.
\begin{proposition}
Consider a semiring $S = \langle A,+,\times,\0,\1 \rangle$, a
domain of the variables $D$, an ordered set of variables $V$, the
corresponding structure $\C$ and the class of hiding functions
$\exists_x : \C \rightarrow \C$ as defined in
Definition~\ref{def:hiding}. Then, for any constraint $c$ and any
variable $x \subseteq V$, $c\Downarrow_{V-x} = \exists_x
c$.
\end{proposition}
\begin{proof}
Is enough to apply the definition of $\Downarrow_{V-x}$ and
$\exists_x$ and check that both are equal to $\sum_{d_i\in D}
c\eta[x := d_i]$.
\end{proof}

We now show how the hiding function so defined satisfies the
properties of cylindric algebras.

\begin{theorem}
Consider a semiring $S = \langle A,+,\times,\0,\1 \rangle$, a
domain of the variables $D$, an ordered set of variables $V$, the
corresponding structure $\C$ and the class of hiding functions
$\exists_x : \C \rightarrow \C$ as defined in
Definition~\ref{def:hiding}. Then $\C$ is a cylindric algebra
satisfying:
\begin{enumerate}
\item $c \vdash \exists_x c$
\item $c_1 \vdash c_2$ implies $\exists_x c_1 \vdash \exists_x c_2$
\item $\exists_x(c_1 \otimes \exists_x c_2) \approx \exists_x c_1 \otimes
\exists_x c_2$,
\item $\exists_x\exists_y c \approx \exists_y\exists_x c$
\end{enumerate}
\end{theorem}
\begin{proof}
Let us consider all the items:
\begin{enumerate}
\item It follows from the intensivity of $+$;
\item It follows from the monotonicity of $+$;
\item It follows from theorems about distributivity and idempotence,
proven in \cite{jacm};
\item It follows from commutativity and associativity of $+$.
\end{enumerate}
\end{proof}

%STEFANO
%versione lunga
% theorema:
% se supp(c) = X allora \Exists_{V-X}c = c

To model parameter passing we need also to define what diagonal
elements are.
\begin{definition}[(diagonal elements)]
\label{def:diag} Consider
%a semiring $S = \langle A,+,\times,\0,\1 \rangle$, a domain of the variables $D$,
an ordered set of variables $V$ and the corresponding soft
constraint system $\C$. Let us define for each $x,y \in V$ a
constraint $d_{xy} \in \C$ s.t., $d_{xy}\eta[x:= a, y := b]= \1$
if $a=b$ and $d_{xy}\eta[x:= a, y := b] = \0$ if $a\neq b$. Notice
that $supp(d_{xy})=\{x,y\}$.
\end{definition}

We can prove that the constraints just defined are diagonal
elements.
% \cite{HMT71,HMT85}.
\begin{theorem}
Consider a semiring $S = \langle A,+,\times,\0,\1 \rangle$, a
domain of the variables $D$, an ordered set of variables $V$, and
the corresponding structure $\C$. The constraints $d_{xy}$ defined
in Definition~\ref{def:diag} represent diagonal elements, that is
\begin{enumerate}
\item $d_{xx} = \1$,
\item if $z \neq x,y$ then $d_{xy} = \exists_z(d_{xz} \otimes d_{zy})$,
\item if $x \neq y$ then $d_{xy} \otimes \exists_x(c \otimes d_{xy}) \ent c$.
\end{enumerate}
\end{theorem}

\begin{proof}

\begin{enumerate}
\item It follows from the definition of the $\1$ constant and of the
diagonal constraint;
\item The constraint $d_{xz} \otimes d_{zy}$ is equal to $\1$ when
$x=y=z$, and is equal to $\0$ in all the other cases. If we
project this constraint over $z$, we obtain the constraint
$\exists_z(d_{xz} \otimes d_{zy})$ that is equal to $\1$ only when
$x=y$;
\item The constraint $(c \otimes d_{xy})\eta$ has value $\0$
whenever $\eta(x)\not = \eta(y)$ and $c\eta$ elsewhere. Now,
$(\exists_x(c \otimes d_{xy}))\eta$ is by definition equal to
$c\eta[x := y]$. Thus $(d_{xy} \otimes
\exists_x(c \otimes d_{xy}))\eta$ is equal to $c\eta$ when
$\eta(x)=\eta(y)$ and $\0$ elsewhere. By the last assumption,
we have the relation of entailment with $c$.
\end{enumerate}
\end{proof}
\subsection{Using cc on top of a soft constraint system}
The only problem in using a soft constraint system in a cc
language is the interpretation of the {\em consistency} notion
necessary to deal with the ask and tell operations.

Usually SCSPs with best level of consistency equal to $\0$ are
interpreted as inconsistent, and those with level greater than
$\0$ as consistent, but we can be more general. In fact, we can
define a suitable function $\alpha$ that, given the best level of
the actual store, will map such a level over the classical notion
of consistency/inconsistency. More precisely, given a semiring $S
= \langle A,+,\times,\0,\1 \rangle$, we can define a function
$\alpha: A \rightarrow \{false, true\}$.
Function $\alpha$ has to be at least monotone, but functions with
a richer set of properties could be used.
It is worth to notice that in a different
environment some of the authors use a similar function to map
elements from a semiring to another, by using abstract
interpretation techniques \cite{abs-sara,abs-cyprus}

Whenever we need to check the consistency of the store, we will
first compute the best level and then we will map such a value by
using function $\alpha$ over $true$ or $false$.

It is important to notice that changing the $\alpha$ function
(that is, by mapping in a different way the set of values $A$ over
the boolean elements $true$ and $false$), the same cc agent yields
different results: by using a high cut level, the cc agent will
either finish with a failure or succeed with a high final best
level of consistency of the store. On the other hand, by using a
low level, more programs will end in a success state.

%%%%

\section{Soft Concurrent Constraint Programming}
\label{sec:scc} The next step in our work is now to extend the
syntax of the language in order to directly handle the cut level.
%When we want to program with soft constraints better than with
%classical constraints, the structure of the language has to take
%in account all the softness notions introduced. First of all in
%the soft framework there are no notions of consistency and
%inconsistency. Instead, a notion of {\em $\alpha$-consistency} is
%introduced.
This means that the syntax and semantics of the tell and ask
agents have to be enriched with a threshold to specify when
tell/ask agents have to fail, succeed or suspend.

Given a soft constraint system $\langle S,D,V\rangle$ and the
corresponding structure $\C$, and any constraint $\phi \in \C$,
the syntax of agents in soft concurrent constraint
programming is given in Table~\ref{tab:scc}.
\begin{small}
\begin{acmtable}{250pt}
%\hrule
%\vskip2pt
\begin{align*}
P ::&= F.A\\
F ::&= p(X)::A \mid F.F\\
A ::&= stop \mid tell(c)
\rightarrow_{\phi} A \mid tell(c) \rightarrow^a A \mid E \mid A \|
A \mid \exists X.A \mid
p(X)\\
E ::&= ask(c)
\rightarrow_{\phi} A \mid ask(c) \rightarrow^a A \mid E+E
\end{align*}
\caption{scc syntax} \label{tab:scc}
%\vskip2pt
%\hrule
\end{acmtable}
\end{small}
The main difference w.r.t. the original {\em cc} syntax is the
presence of a semiring element $a$ and of a constraint $\phi$
to be checked whenever an {\em
ask} or {\em tell} operation is performed. More precisely, the
level $a$ (resp., $\phi$) will be used as a cut level to prune computations
that are not good enough.

%\subsection{Operational Semantics}
\label{sec:semop} We present here a structured operational
semantics for scc programs, in the SOS style, which consists of
defining the semantic of the programming language by specifying a
set of {\em configurations} $\Gamma$, which define the states
during execution, a relation $\rightarrow\ \subseteq
\Gamma\times\Gamma$ which describes the {\em transition} relation
between the configurations, and a set $T$ of {\em terminal}
configurations.
To give an operational semantics to
our language, we need to describe an appropriate transition system.

\begin{definition}[(transition system)]
A transition system is a triple
\mbox{$\langle \Gamma, T, \rightarrow \rangle$}
where $\Gamma$ is a set of possible configurations, $T
\subseteq \Gamma $ is the set of {\em terminal} configurations and
$\rightarrow \subseteq \Gamma \times \Gamma$ is a binary relation
between configurations.
\end{definition}

The set of configurations represent the evolutions of the agents
and the modifications in the constraint store. We define the
transition system of soft cc as follows:
\begin{definition}[(configurations)]
The set of configurations for a soft cc system is the set $\Gamma
= \{\langle A, \sigma\rangle\} \cup \{\langle success, \sigma
\rangle\}$, where $\sigma \in \mathcal{C}$. The set of terminal
configurations is the set $T=\{\langle success, \sigma \rangle\}$
and the transition rule for the scc language are defined in
Table~\ref{tab:transition}.
\end{definition}
\begin{small}
\begin{acmtable}{250pt}
%\hrule %\vskip2pt
\begin{gather*}
\trans{\langle stop, \sigma \rangle}{} {\langle success, \sigma
\rangle} \tag{Stop}
\\
\irule{(\sigma \otimes c)\Downarrow_{\emptyset} \not<
a}{\trans{\langle tell(c) \rightarrow^{a} A, \sigma \rangle} {}
{\langle A, \sigma \otimes c\rangle}}\tag{Valued-tell}
\\
\irule{\sigma \otimes c \not\sqsubset \phi }{\trans{\langle
tell(c) \rightarrow_{\phi} A, \sigma \rangle} {} {\langle A,
\sigma \otimes c\rangle}}\tag{Tell}
\\
\irule{\sigma \ent c, \sigma\Downarrow_{\emptyset} \not< a}
{\trans{\langle ask(c) \rightarrow^a A, \sigma \rangle} {}
{\langle A, \sigma \rangle}}\tag{Valued-ask}
\\
\irule{\sigma \ent c, \sigma \not\sqsubset \phi} {\trans{\langle
ask(c) \rightarrow_{\phi} A, \sigma \rangle} {} {\langle A, \sigma
\rangle}}\tag{Ask}
\\
\irule{\trans{\langle A_1,\sigma \rangle}{}{\langle A_1',\sigma'
\rangle}}{\begin{matrix}\trans{\langle A_1 \| A_2, \sigma \rangle}
{} {\langle A_1' \| A_2, \sigma' \rangle}\\
\trans{\langle A_2 \| A_1, \sigma \rangle} {} {\langle A_2 \|
A_1', \sigma' \rangle}\end{matrix}} \qquad \irule{\trans{\langle
A_1,\sigma \rangle}{}{\langle success,\sigma'
\rangle}}{\begin{matrix}\trans{\langle A_1 \| A_2, \sigma \rangle}
{} {\langle A_2, \sigma' \rangle}\\
\trans{\langle A_2 \| A_1, \sigma \rangle} {} {\langle A_2,
\sigma' \rangle}\end{matrix}} \tag{Parallelism}
\\
\irule{\trans{\langle E_1,\sigma \rangle}{}{\langle A_1,\sigma'
\rangle}} {\begin{matrix}\trans{\langle E_1 + E_2, \sigma \rangle}
{} {\langle A_1, \sigma' \rangle}\\\trans{\langle E_2 + E_1,
\sigma \rangle} {} {\langle A_1, \sigma'
\rangle}\end{matrix}}\tag{Nondeterminism}
\\
\irule{\trans{\langle A[y/x], \sigma\rangle}{}{\langle
A',\sigma'\rangle}} {\trans{\langle \exists_x A, \sigma \rangle}
{} {\langle A', \sigma'\rangle}}\text{ with $y$ {\em
fresh}}\tag{Hidden variables}
\\
\trans{\langle p(y), \sigma\rangle} {} {\langle A[y/x], \sigma
\rangle} \text{ when $p(x) :: A$}\tag{Procedure call}
\end{gather*}
\caption{Transition rules for scc}
\label{tab:transition}
%\vskip2pt
%\hrule
\end{acmtable}
\end{small}
Here is a brief description of the transition rules:
\begin{description}
  \item[Stop] The stop agent succeeds in one step by transforming
  itself into terminal configuration {\em success}.
  \item[Valued-tell] The valued-tell rule checks for the $\alpha$-consistency
  of the SCSP defined by the store $\sigma \otimes c$. The rule can
  be applied only if the store $\sigma \otimes c$ is $b$-consistent
  with $b \not< a$. In this case the agent evolves to the new
  agent $A$ over the store $\sigma \otimes c$.
  Note that different
  choices of the {\em cut level} $a$ could possibly lead to different
  computations.
  \item[Tell] The tell action is a finer check of the
  store. In this case, a pointwise comparison between the store
  $\sigma \otimes c$ and the constraint $\phi$
  is performed. The idea is to perform an overall
  check of the store and to continue the computation only if there is
  the possibility to compute a solution not worse
  than $\phi$.
  \item[Valued-ask] The semantics of the valued-ask is extended in
  a way similar to what we have done for the valued-tell action. This
  means that, to apply the rule, we need to check if the store $\sigma$
  entails the constraint $c$ and also if the store is ``consistent
  enough'' w.r.t. the threshold $a$ set by the programmer.
  \item[Ask] Similar to the {\em tell} rule, here a finer (pointwise)
  threshold $\phi$ is compared to the store $\sigma$.
  \item[Nondeterminism and parallelism] The composition
  operators $+$ and $\|$ are not modified w.r.t. the classical
  ones: a parallel agent will succeed if all the agents succeeds;
  a nondeterministic rule chooses any agent whose guard succeeds.
  %Note that at this stage no distinction between the {\em
  %don't know} or {\em don't care} observational behaviours is
  %made.
  \item[Hidden variables] The semantics of the existential
  quantifier is similar to that described in \cite{vijay-book} by
  using the notion of {\em freshness} of the new variable added to
  the store.
% A different way to describe this could be by
%  using the approach in
%  \cite{BP94,JSS91,BPP95,PW98,PWcats98,BoerGMP97,VIJAY-BOOK} by
%  substituting the agent $\exists_x A$ with $\exists_x^{\1} A$
%  and by using the two rules:
%  \begin{gather*}
%\irule{\trans{\langle A, c \otimes \exists_x \sigma
%\rangle}{}{\langle A',c'\rangle}} {\trans{\langle \exists_x^c A,
%\sigma \rangle} {} {\langle A', \sigma \otimes \exists_x^{c'}
%\rangle}} \qquad \irule{\trans{\langle A, c \otimes \exists_x
%\sigma \rangle}{}{\langle stop, c'\rangle}} {\trans{\langle
%\exists_x^c A, \sigma \rangle} {} {\langle stop, \sigma \otimes
%\exists_x^{c'} \rangle}}
%\end{gather*}
%In the agent $\exists_x^c A$, $c$ is a local store of $A$
%containing information on $x$ which is hidden in the external
%store. Initially the local store is empty and so $\exists_x A$ is
%substituted by $\exists_x^{\1} A$.
\item[Procedure calls] The semantics of the procedure call is not modified
w.r.t. the classical one. The only difference is the different use
of the diagonal constraint to represent parameter passing.
\end{description}

\subsection{Eventual Tell/Ask}

We recall that both ask and tell operations in cc could be either atomic
(that is, if the corresponding check is not satisfied, the agent
does not evolve) or eventual (that is, the agent evolves regardless of the
result of the check).
It is interesting to notice that
the transition rules defined in Table~\ref{tab:transition} could
be used to provide both interpretations of the ask and tell
operations. In
fact, while the generic tell/ask rule represents an atomic
behaviour, by setting $\phi = \0$ or $a=\0$ we obtain their {\em
eventual} version:
\begin{gather*}
\trans{\langle tell(c) \rightarrow A, \sigma \rangle} {} {\langle
A, \sigma \otimes c\rangle}\tag{Eventual tell}
\\
\irule{\sigma \ent c} {\trans{\langle ask(c) \rightarrow A, \sigma
\rangle} {} {\langle A, \sigma \rangle}}\tag{Eventual ask}
\end{gather*}
Notice that, by using an eventual interpretation, the transition
rules of the scc become the same as those of cc (with an eventual
interpretation too). This happens since, in the eventual version,
the tell/ask agent never checks for consistency and so the soft
notion of $\alpha$-consistency does not play any role.

\section{A Simple Example}
\label{sec:example}
In this section we will show the behaviour of
some of the rules of our transition system. We consider in this
example a soft constraint system over the fuzzy semiring. Consider
the fuzzy constraints
\begin{align*}
c:\{x,y\} \rightarrow {\mathcal R}^2
\rightarrow [0,1] \hspace{1cm} &\text{ s.t. }
c(x,y) =\frac{1}{1+|x-y|} \hspace{1cm} \text{
and } \\
c':\{x\} \rightarrow {\mathcal R}
\rightarrow [0,1] \hspace{1cm} &\text{ s.t.  }
c'(x) =
\begin{cases}
1 & \text{if $x\leq 10$},\\
0 & \text{otherwise}.\\
\end{cases}
\end{align*}
Notice that the domain of both variables $x$ and $y$ is in this example any
integer (or real) number.
As any fuzzy CSP, the definition of the constraints is instead in the interval
$[0,1]$.

Let's now evaluate the agent
$$\langle tell(c)\rightarrow^{0.4} ask(c')
\rightarrow^{0.8} stop, 1 \rangle$$
in the empty starting store
$1$.

By applying the {\em Valued-tell} rule we need to check $(1
\otimes c)\Downarrow_{\emptyset}\not<0.4$. Since $1 \otimes c =
c$ and $c\Downarrow_{\emptyset} = 1$, the agent can perform the
step, and it reaches the state
$$\langle ask(c') \rightarrow^{0.8}
stop, c\rangle.$$
Now we need to check (by following the rule of
{\em Valued-ask}) if $c \ent c'$ and
$c\Downarrow_{\emptyset}\not<0.8$. While the second relation
easily holds, the first one does not hold (in fact, for $x=11$ and
$y=10$ we have $c'(x)=0$ and $c(x,y)=0.5$).

If instead we consider
the constraint
%\[c":\{x,y\} \rightarrow {\mathcal R}
%\rightarrow [0,1] \text{ s.t. }
$c''(x,y) =\frac{1}{1 + 2 \times |x-y|}$
in place of $c'$, then we have
$$\langle ask(c'') \rightarrow^{0.8}
stop, c\rangle.$$
Here
the condition $c \ent c"$ easily holds and
the agent $ask(c'') \rightarrow^{0.8} stop$ can perform its last
step, reaching the $stop$ and $success$ states:
$$\langle stop, c
\otimes c'' \rangle \rightarrow \langle success, c \otimes
c'' \rangle.$$

\section{Observables and Cuts}
Sometimes one could desire to see an agent, and a
corresponding program, execute with a cut level
which is different from the one originally given.
We will therefore define $cut_{\psi}(A)$ the agent $A$
where all the occurrences of any cut level, say $\phi$,
in any subagent of $A$ or in any clause of the program,
are replaced by $\psi$ if $\phi \sqsubseteq \psi$. This means that
the cut level of each subagent and clause becomes
at least $\psi$, or is left to the original level.

In this paper, for simplicity and generality
reasons, this cut level change applies
%, for this paper,
only to those programs with cut levels which are constraints ($\phi$),
and not single semiring levels ($a$).

\begin{definition}[(cut function)]
\label{def:cutfunct} Consider an scc agent $A$; we define the
function $cut_{\psi}:A\rightarrow A$ that transforms ask and tell
subagents as follows:
\[
cut_{\psi}(ask/tell(c)\rightarrow_{\phi}) =
\begin{cases}
ask/tell(c)\rightarrow_{\psi} & \text{if $\phi \sqsubset \psi$},\\
ask/tell(c)\rightarrow_{\phi} & \text{otherwise}.\\
\end{cases}
\]
\end{definition}
By definition of $cut_{\psi}$, it is easy to see that $cut_{\0}(A)=A$.

\Commento{
Non ho modificato la funzione di taglio perche nell'altro modo:
\[
cut_{\psi}(ask/tell(c)\rightarrow_{\phi}) =
\begin{cases}
ask/tell(c)\rightarrow_{\phi} & \text{if $\phi \sqsupset \psi$},\\
ask/tell(c)\rightarrow_{\psi} & \text{otherwise}.\\
\end{cases}
\]
Posso introdurre con un taglio computazioni nuove che non vorrei:
supponi infatti di avere $langle tell(1)\rightarrow_{phi} A, \sigma \rangle$.
Se applico il taglio $\psi$ con $\psi$ inconfrontabile con $\phi$,
nel vecchio caso non modifico l'agente (e quindi sicuramente non
aggiungo computazioni).

Nel nuovo caso potrei aver avuto $\sigma \sqsubset \phi$ e quindi una computazione che falliva.
Dopo il taglio potrei avere $\sigma$ inconfrontabile con $\psi$, e quindi,
il passo potrebbe avere successo.

Avrei quindi aggiunto una computazione.
}

We can then prove the following Lemma (that will be useful later):

\begin{lemma}[(tell and ask cut)]
\label{ask-tell-cut}
Consider the Tell and Ask rules of Table \ref{tab:transition},
and the constraints
$\sigma$ and $c$ as defined in such rules. Then:
\begin{itemize}
\item
If the {\em Tell} rule can be applied to agent $A$, then the rule
can be applied also to $cut_{\psi}(A)$ when when ${\psi}\sqsubseteq
\sigma\otimes c$.
\item
If the {\em Ask} rule can be applied to agent $A$, then the rule
can be applied also to $cut_{\psi}(A)$ when ${\psi}\sqsubseteq \sigma$.
\end{itemize}
\end{lemma}
\begin{proof}
We will prove only the first item; the second can be easily
proved by using the same ideas.
By the definition of the tell transition rules of
Table~\ref{tab:transition}, if we can apply the rule it means that
$A::=tell(c) \rightarrow_{\phi} A'$ and if $\sigma$ is the store
we have $\sigma \otimes c \not\sqsubset \phi$.
Now, by definition of $cut_{\psi}$, we can have
\begin{itemize}
\item $cut_{\psi}(A)::=tell(c) \rightarrow_{\psi} cut_{\psi}(A')$
when $\phi \sqsubset \psi$.
\item $cut_{\psi}(A)::=tell(c) \rightarrow_{\phi} cut_{\psi}(A')$
when $\phi \not \sqsubset \psi$,
\end{itemize}
In the first case, the statement holds by initial hypothesis over $A$.
In the second case, since by hypothesis we have $\sigma \otimes c \not\sqsubset
{\psi}$, again the statement holds by the definition of the tell transition
rules of Table~\ref{tab:transition}.
\end{proof}

It is now interesting to notice that the thresholds appearing in
the program are related to the final computed stores:

\begin{theorem}[(thresholds)]
\label{threshold2} Consider an scc computation
\[\langle A,\1
\rangle \rightarrow \langle A_1, \sigma_1 \rangle \rightarrow
\ldots \langle A_n, \sigma_n \rangle \rightarrow \langle success,
\sigma \rangle\]
 for a program $P$. Then, also
 \[\langle
cut_{\sigma}(A),\1 \rangle \rightarrow \langle cut_{\sigma}(A_1),
\sigma_1 \rangle \rightarrow \ldots \langle cut_{\sigma}(A_n),
\sigma_n \rangle \rightarrow \langle success, \sigma \rangle\]
 is
an scc computation for program $P$.
\end{theorem}
\begin{proof}
First of all, notice that during the computation an agent
can only add constraints to the store. So, since $\times$ is extensive,
the store can only
monotonically decrease starting from the initial store $\1$
and ending in the final store $\sigma$.
So we have
\[
\1 \sqsupseteq \sigma_1 \ldots \sqsupseteq \sigma_n \sqsupseteq \sigma.
\]
Now, the statement follows by applying at each step the results of
Lemma~\ref{ask-tell-cut}.
In fact, at each step the hypothesis of the lemma hold:
\begin{itemize}
\item the cut $\sigma$ is always lower than the current store
(${\sigma}\sqsubseteq
\sigma_i\otimes c$);
\item the ask and tell operations can be applied (moving from agent
$A_i$ to agent $A_i+1$).
\end{itemize}
%
%Since at each step the
%store is never lower than $\sigma$, by~Lemma \ref{ask-tell-cut}
%the theorem easily holds.
\end{proof}

%%%
\subsection{Capturing Success Computations.}
\label{sec:success}
Given the transition system as defined in the previous section, we
now define what we want to observe of the program behaviour
as described by the transitions. To do this, we define for each agent
$A$ the set of constraints
\[
\mathcal{S}_A = \{\sigma\Downarrow_{var(A)} \mid \langle A,\1
\rangle \rightarrow^* \langle success, \sigma \rangle\}
\]
that collects the results of the successful computations that the
agent can perform. Notice that the computed store $\sigma$ is projected over
the variables of the agent $A$ to discard any {\em fresh} variable
introduced in the store by the $\exists$ operator.
%In this paper
%we only consider a semantics that collects success states. We plan
%to extend the operational semantics to collect also failing and
%suspending computations.

The observable $\mathcal{S}_A$ could be refined by considering,
instead of the set of successful computations starting from
$\langle A,\1 \rangle$, only a subset of them. For example,
one could be
interested in considering only the {\em best}
computations: in this case, all the computations leading to a
store worse than one already collected are disregarded. With a
pessimistic view, the representative subset could instead collect
all the worst computations (that is, all the computations better
than others are disregarded). Finally, also a set containing both
the best and the worst computations could be considered. These
options are reminiscent of Hoare, Smith and Egli-Milner
powerdomains respectively \cite{powerdomains}.

%Let us also notice that different cut levels in the ask and tell
%operations could lead to a different final sets $\mathcal{S}_A$.
%In fact, it can be proved that if the thresholds of the ask and
%tell operations of the program are not worse than a given
%$\alpha$, we can be sure to find in the final store only solutions
%not worse than $\alpha$. This observation can be useful when we
%are looking just for the best stores reachable from an initial
%given agent. In fact, we can move the cut up and down (in a way
%similar to a binary search) and perform a branch and bound
%exploration of the search tree in order to find the final success
%sets.

%STEFANO
%versione lunga:
% add citations a hoare, milner, dijstra

At this stage, the difference between don't know and don't care
nondeterminism
%\footnote{The notion of don't know commitment in the
%context of concurrent languages, was presented in \cite{sar85c}
%and its use in concurrent constraint languages was explained in
%\cite{vijay-book}.\\ The notion of don't care commitment is a
%generalization of Prolog cut operation and was apparently
%introduced to the field of definite clause based programming
%languages in \cite{JLOGP::ClarkG1985}. The use of don't care
%committed choice (together the don't know committed choice) in the
%cc framework was introduced and discussed in \cite{sar85c,vijay-book}.}
%(called sometimes also angelic/demonic nondeterminism or
%nondeterminism/indeterminism or nondeterminate/committed choice)
arises only in the way the {\em observables} are interpreted: in a
don't care approach, agent $A$ can commit
%indeterministically
to {\em one} of the final stores $\sigma\Downarrow_{var(A)}$,
while, in a don't know approach, in classical cc programming it is
enough that one of the final stores is consistent. Since
existential quantification corresponds to the sum in our
semiring-based  approach, for us a don't know approach leads to
the sum (that is, the lub) of all final stores:
%the stores collected in $\mathcal{S}_A$:
\[
\mathcal{S}^{dk}_A = \bigoplus_{\sigma \in \mathcal{S}_A} \sigma.
\]

It is now interesting to notice that the thresholds appearing in
the program are related also to the observable sets:

\begin{proposition}[(Thresholds and $\mathcal{S}_A$ (1))]
\label{threshold3adj}
For each $\psi$, we have
$\mathcal{S}_A \supseteq \mathcal{S}_{cut_{\psi}(A)}$.
\end{proposition}
\begin{proof}
By definition of cuts (Definition~\ref{def:cutfunct}), we can
modify the agents only by changing the thresholds with a new level,
greater than the previous one.
So, easily, we can only cut away some computations.
\end{proof}

\begin{corollary}[(Thresholds and $\mathcal{S}^{dk}_A$ (1))]
\label{threshold3adj2}
For each $\psi$, we have
$\mathcal{S}^{dk}_A \supseteq \mathcal{S}^{dk}_{cut_{\psi}(A)}$.
\end{corollary}
\begin{proof}
It follows from the definition of $\mathcal{S}^{dk}_A$ and from
Proposition~\ref{threshold3adj}.
\end{proof}

\begin{theorem}[(Thresholds and $\mathcal{S}_A$ (2))]
\label{threshold3} Let $\psi \sqsubseteq glb\{\sigma \in
\mathcal{S}_A\}$. Then $\mathcal{S}_A =
\mathcal{S}_{cut_{\psi}(A)}$.
\end{theorem}
\begin{proof}
By Proposition~\ref{threshold3adj}, we have $\mathcal{S}_A
\subseteq \mathcal{S}_{cut_{\psi}(A)}$.
Moreover, since $\psi$ is lower than all $\sigma$ in $\mathcal{S}_A$, by
Theorem~\ref{threshold2} we have that all the computations are also in
$\mathcal{S}_{cut_{\psi}(A)}$.
So, the statement follows.
\end{proof}

Notice that, thanks to Theorem~\ref{threshold3} and to
Proposition~\ref{threshold3adj}, whenever we
have a lower bound $\psi$ of the glb of the final solutions,
we can use $\psi$ as a threshold to eliminate some computations.
Moreover, we can prove the following theorem:

\begin{theorem}
\label{theo:poker}
Let $\sigma \in \mathcal{S}_A$ and $\sigma \not\in
\mathcal{S}_{cut_{\psi}(A)}$. Then we have $\sigma \sqsubset \psi$.
\end{theorem}
\begin{proof}
If $\sigma \in \mathcal{S}_A$ and $\sigma \not\in
\mathcal{S}_{cut_{\psi}(A)}$, it means that the cut eliminates some computations.
So, at some step we have changed the threshold of some tell or ask agent.
In particular, since we know by Theorem~\ref{threshold2} that when
$\psi \sqsubseteq \sigma$ we do not modify the computation,
we need $\psi \not \sqsubseteq \sigma$. Moreover,
since the tell and ask rules fail only if $\sigma \sqsubset \psi$,
we easily have the statement of the theorem.
\end{proof}

The following theorem relates thresholds and $\mathcal{S}^{dk}_A$.
\begin{theorem}[(Thresholds and $\mathcal{S}^{dk}_A$ (2))]
\label{theo:threshdk}
Let $\Psi_A = \{\sigma \in \mathcal{S}_A \mid \not\exists
\sigma'\in \mathcal{S}_A \text{ with } \sigma'\sqsupseteq
\sigma\}$ (that is, $\Psi_A$ is the set of ``greatest'' elements
of $\mathcal{S}_A$). Let also $\psi \sqsubseteq glb\{\sigma \in
\Psi_A\}$. Then $\mathcal{S}^{dk}_A =
\mathcal{S}^{dk}_{cut_{\psi}(A)}$.
\end{theorem}
\begin{proof}
%Easily follows by Theorem~\ref{threshold2} and by the
%associativity and distributivity of $\otimes$ and $\oplus$.
%By Theorem~\ref{threshold3} we easily have that
%$\Psi_A = \mathcal{S}_{cut_{\psi}(A)}$.
%
%Let now consider $\mathcal{S}^{dk}_A$. Since when $\sigma_1
%\sqsubseteq \sigma_2$ we have $\sigma_1 \oplus \sigma_2 =
%\sigma_2$; then $\mathcal{S}^{dk}_A =\bigoplus_{\sigma \in
%\mathcal{S}_A} \sigma$ is also equal to $\bigoplus_{\sigma \in
%\Psi_A} \sigma$ (in fact, in $\mathcal{S}_A - \Psi_A$ we have only
%elements strictly lower than those present in $\Psi_A$).
%
%Now, since $\Psi_A = \mathcal{S}_{cut_{\psi}(A)}$ we obtain the
%thesis.
%seconda versione
%Since when $\sigma_1 \sqsubseteq \sigma_2$ we have $\sigma_1
%+\sigma_2 = \sigma_2$, we can eliminate from $\mathcal{S}_A$ all
%the $\sigma: \exists \sigma'\in \mathcal{S}_A$ with $\sigma
%\sqsubseteq \sigma'$. By hypothesis, this set is just $\Psi_A$. So
%easily follows $\mathcal{S}^{dk}_A =\bigoplus_{\sigma \in \Psi_A}
%\sigma$; now, by Theorem~\ref{threshold3}, we have $\{\sigma\in
%\Psi_A \}= \mathcal{S}_{cut_{\Psi}(A)}$ and so, also
%$\mathcal{S}^{dk}_A = \mathcal{S}^{dk}_{cut_{\psi}(A)}$.
Since we have $a+b=b \iff a \leq b$, we easily have
$\bigoplus_{\sigma \in \mathcal{S}_A} \sigma =
\bigoplus_{\sigma \in \Psi_A} \sigma$.
Now, by following a reasoning similar to Theorem~\ref{threshold3}, by applying
a cut with a threshold $\psi \sqsubseteq glb\{\sigma \in \Psi_A\}$ we do not eliminate
any computation.
So we obtain $\mathcal{S}^{dk}_A =
\bigoplus_{\sigma \in \mathcal{S}_A} \sigma =
\bigoplus_{\sigma \in \Psi_A} \sigma =
\mathcal{S}^{dk}_{cut_{\psi}(A)}$
\end{proof}

\begin{lemma}
\label{lemma:tris}
Given any constraint $\psi$, we have:
\[
\mathcal{S}^{dk}_A
\sqsubseteq
\psi + \mathcal{S}^{dk}_{cut_{\psi}(A)}.
\]
\end{lemma}
\begin{proof}
Let $S$ be the set of all solutions; then $\mathcal{S}^{dk}_A =
lub(S)$ and $\mathcal{S}^{dk}_{cut_{\psi}(A)} = lub(S_1)$ where
$S_1 \subseteq S$. The solutions that have been eliminated by the cut
$\psi$ (that is all the $\sigma \in S - S_1$) are all lower than
$\psi$ by Theorem~\ref{theo:poker}.
So, it easily follows that $\mathcal{S}^{dk}_A
\sqsubseteq
\psi + \mathcal{S}^{dk}_{cut_{\psi}(A)}$.
\end{proof}

\begin{theorem}
\label{theo:psi}
Given any constraint $\psi$, we have:
\[
\mathcal{S}^{dk}_{cut_{\psi}(A)} \sqsubseteq \mathcal{S}^{dk}_A
\sqsubseteq \psi + \mathcal{S}^{dk}_{cut_{\psi}(A)}.
\]
\end{theorem}
\begin{proof}
From Corollary~\ref{threshold3adj2}, we have
$\mathcal{S}^{dk}_{cut_{\psi}(A)} \sqsubseteq \mathcal{S}^{dk}_A$.
From Lemma~\ref{lemma:tris} we have instead
$\mathcal{S}^{dk}_A
\sqsubseteq \psi + \mathcal{S}^{dk}_{cut_{\psi}(A)}$.
\end{proof}

This theorem suggests a way to cut useless computations while
generating the observable $\mathcal{S}^{dk}_A$
of an scc program $P$ starting from agent $A$.
A very naive way to obtain such an observable would be to first generate
all final states, of the form $\langle success, \sigma_i \rangle$,
and then compute their lub.
An alternative, smarter way to compute this same observable would be
to do the following.  First we start executing the program
as it is, and find
a first solution, say $\sigma_1$. Then we restart the execution
applying the cut level $\sigma_1$.

By Theorem~\ref{theo:threshdk}, this new cut level
cannot eliminate solutions which influence the computation of the
observable: the only solutions it will cut are those
that are lower than the one we already found, thus
useless in terms of the computation of $\mathcal{S}^{dk}_A$.

In general, after having found solutions $\sigma_1, \ldots,
\sigma_k$, we restart execution with cut level $\psi =
\sigma_1 + \ldots + \sigma_k$.
Again, this will not cut
crucial solutions but only some that are lower
than the sum of those already found.
When the execution of the program
terminates with no solution
% or with a new
%cut level with coincides with the previous one,
we can be sure that the cut
level just used (which is the sum of all solutions found) is the
desired observable (in fact, by Theorem~\ref{theo:psi} when
$\mathcal{S}^{dk}_{cut_{\psi}(A)} = \psi$ we necessarily have
$\mathcal{S}^{dk}_{cut_{\psi}(A)} = \mathcal{S}^{dk}_A
= \psi$).

In a way, such an execution method
resembles a branch \& bound strategy, where the cut levels
have the role of the bounds.

%\item If $\mathcal{S}^{dk}_{cut_{\psi}(A)} =
%\mathcal{S}^{dk}_{cut_{\psi}(A)} + \psi$,
%it means that we have obtained the observable of $P$.
%So we can stop.

%\item If instead $\mathcal{S}^{dk}_{cut_{\psi}(A)} <
%\mathcal{S}^{dk}_{cut_{\psi}(A)} + \psi$,
%it means that the observable of $P$ could be larger than what we
%have obtained, so we ask the oracle for a lower threshold
%(if it exists), and
%restart from step 1 with the new threshold.
The following corollary is important to show
the correctness of this approach.

\begin{corollary}
\label{theo:psi2}
Given any constraint $\psi \sqsubseteq \mathcal{S}^{dk}_A$, we have:
\[
\mathcal{S}^{dk}_A
= \psi + \mathcal{S}^{dk}_{cut_{\psi}(A)}.
\]
\end{corollary}
\begin{proof}
It easily comes from Theorem~\ref{theo:psi}.
\end{proof}

%\subsubsection{Correctness of the algorithm}
Let us now use this corollary to prove the correctness of the
whole procedure above.

Let $\sigma_1$ be the first final state reached by agent $A$. By
stopping the algorithm after one step, what we have to prove is
$\mathcal{S}^{dk}_A = \sigma_1 +
\mathcal{S}^{dk}_{cut_{\sigma_1}(A)}$. Since $\sigma_1$ is
for sure lower than $\mathcal{S}^{dk}_A$, this is true by
Corollary~\ref{theo:psi2}.

By applying this procedure iteratively, we will collect a superset
$\Psi'_A$ of $\Psi_A = \{\sigma \in \mathcal{S}_A \mid \not\exists
\sigma'\in \mathcal{S}_A \text{ with } \sigma'\sqsupseteq
\sigma\}$ ($\Psi'_A$ is a superset of $\Psi_A$ because we could
collect a final state $\sigma_i$ before computing a final state
$\sigma_j \sqsupseteq \sigma_i$; in this case both will be in
$\Psi'_A$). Even if $\Psi'_A$ contains more elements then
$\Psi_A$, we have $\bigoplus_{\sigma \in \Psi'_A} =
\bigoplus_{\sigma \in \Psi_A}$ (for the extensivity and idempotence
properties of $+$).

The only difference with the procedure we have tested correct
w.r.t. the algorithm is that, at each step, it performs a cut by
using the sum of all the previously computed final state.
This means that the algorithm can at each step eliminate
more computations,
but by the results of Theorem~\ref{theo:poker}
the eliminated computations does not change the final result.
%
%We claim
%the two procedure obtain the same result. In fact, we can prove
%$\mathcal{S}^{dk}_{cut_{\sigma_2}(cut_{\sigma_1}(A))} =
%\mathcal{S}^{dk}_{cut_{\sigma_1+ \sigma_2}(A)}$.
%
%\begin{theorem}
%Given any final state  $\sigma_1$ and $\sigma_2$, we have:
%$\mathcal{S}^{dk}_{cut_{\sigma_2}(cut_{\sigma_1}(A))} =
%\mathcal{S}^{dk}_{cut_{\sigma_1+ \sigma_2}(A)}$.
%\end{theorem}
%\begin{proof}
%By definition, $cut_{\sigma_2}(cut_{\sigma_1}(A))$ eliminate all
%the computations of $A$, whose final state is lower than
%$\sigma_2$ and after all the computations leading to a final state
%lower than $\sigma_1$. On the other side, $cut_{\sigma_1+
%\sigma_2}(A)$ eliminate all the final state lower than both
%$\sigma_1$ and $\sigma_2$. Since by definition of $\oplus$, there
%are no $\sigma$ s.t. $\sigma_1 \sqsubseteq \sigma \sqsubseteq
%\sigma_1 \oplus \sigma_2$ and $\sigma_2 \sqsubseteq \sigma
%\sqsubseteq \sigma_1 \oplus \sigma_2$, then the result of the two
%computation is the same.
%\end{proof}

\subsection{Failure}
\label{sec:fail}
The transition system we have defined considers only successful
computations. If this could be a reasonable choice in a don't know
interpretation of the language it will lead to an insufficient
analysis of the behaviour in a {\em pessimistic} interpretation of
the indeterminism. To capture agents' failure, we add
the terminal {\em fail} to the configurations
and the transition rules of
Table~\ref{tab:tr-fail} to those of Table~\ref{tab:transition}.
\begin{acmtable}{250pt}
%\hrule
%\vskip2pt
\begin{gather*}
\irule{\sigma \otimes c \sqsubset \phi}{\trans{\langle tell(c)
\rightarrow_{\phi} A, \sigma \rangle} {} {fail}}\tag{Tell$_1$}
\\
\irule{(\sigma \otimes c)\Downarrow_{\emptyset} <
a}{\trans{\langle tell(c) \rightarrow^a A, \sigma \rangle} {}
{fail}}\tag{Valued-tell$_1$}
\\
\irule{\sigma \sqsubset \phi} {\trans{\langle ask(c)
\rightarrow_{\phi} A, \sigma \rangle} {} {fail}} \tag{Ask$_1$}
\\
\irule{\sigma\Downarrow_{\emptyset} < a} {\trans{\langle ask(c)
\rightarrow^a A, \sigma \rangle} {} {fail}}\tag{Valued-ask$_1$}
\\
\irule{\trans{\langle E_1,\sigma \rangle}{}{fail}, \trans{\langle
E_2,\sigma \rangle}{}{fail}} {\begin{matrix}\trans{\langle E_1 +
E_2, \sigma \rangle} {} {fail}\\ \trans{\langle E_2 + E_1, \sigma
\rangle} {} {fail}\end{matrix}} \tag{Nondeterminism$_1$}
\\
\irule{\trans{\langle A_1,\sigma
\rangle}{}{fail}}{\begin{matrix}\trans{\langle A_1 \| A_2, \sigma
\rangle} {} {fail}\\ \trans{\langle A_2 \| A_1, \sigma \rangle} {}
{fail}\end{matrix}} \tag{Parallelism$_1$}
\end{gather*}
\caption{Failure in the scc language}
\label{tab:tr-fail}
%\vskip2pt \hrule
\end{acmtable}

\begin{description}
  \item[(Valued)tell$_1$/ask$_1$] The failing rule for ask and tell simply
  checks if the added/checked constraint $c$ is {\em inconsistent} with
  the store $\sigma$ and in this case stops the computation and
  gives {\em fail} as a result. Note that since we use soft
  constraints we enriched this operator with a threshold ($a$ or $\phi$).
  This is used also to compute failure. If the level of
  consistency of the resulting store is lower than the threshold level,
  then this is considered a failure.
  \item[Nondeterminism$_1$] Since the failure of a branch arises
  only from the failure of a guard, and since we use angelic non-determinism
  (that is, we check the guards before choosing one path), we
  fail only when all the branches fail.
  \item[Parallelism$_1$] In this case the computation fails as soon as
  one of the branches ails.
%  \item[nota1] se vogliamo invece di usare $fail$ potremmo usare
%  $\langle \sigma, fail \rangle$ per catturale anche lo store del
%  fallimento .. ma non so se abbia un significato (puo'avere solo
%  significato in fase di test se poi l'utente riesegue il
%  programma con una $\phi$ sulle frecce piu'bassa.
\end{description}

The observables of each agent can now be enlarged by using the function
\[
\mathcal{F}_A = \{fail \mid \langle A,\1_V \rangle \rightarrow^*
fail\}
\]
that computes a failure if at least a computation of agent $A$ fails.

By considering also the failing computations, the difference
between don't know and don't care becomes finer.
In fact, in situations where we have $\mathcal{S}_A =
\mathcal{S}^{dk}_A$, the failing computations could make the
difference: in the don't care approach the notion of failure is
{\em existential} and in the don't know one becomes {\em
universal} \cite{BP94}:
\[
\mathcal{F}^{dk}_A = \{fail \mid \text{all computations for $A$
which lead to $fail$}\}.
\]
This means that in the don't know nondeterminism we are interested
in observing a failure only if all the branches fail. In this way,
given an agent $A$ with an empty $\mathcal{S}^{dk}_A$ and a
non-empty $\mathcal{F}^{dk}_A$, we cannot say for sure that the
semantic of this agent is $fail$. In fact, the transition rules we
have defined do not consider {\em hang} and infinite computations.
Similar semidecibility results for soft constraint logic
programming are proven in \cite{toplas00}.

\subsection{Hanging and infinite computations}
\label{sec:hang}
To complete the possible observables of a goal, we need also to
observe the hanged states or those representing infinite
computation. To this extent,
we extend the configurations with the terminals {\em
hang} and $\bot$, and we add some transition rules
(those in Table~\ref{tab:tr-hang}) to handle hanging computations.
\begin{acmtable}{250pt}
%\hrule \vskip2pt
\begin{gather*}
\irule{\sigma \not\ent c, \sigma \not\sqsubset \phi}
{\trans{\langle ask(c) \rightarrow_{\phi} A, \sigma \rangle} {}
{hang}} \tag{Ask$_2$}
\\
\irule{\sigma \not\ent c, \sigma\Downarrow_{\emptyset} \not< a}
{\trans{\langle ask(c) \rightarrow^a A, \sigma \rangle} {}
{hang}}\tag{Valued-ask$_2$}
\\
\irule{\trans{\langle E_1,\sigma \rangle}{}{fail/hang},
\trans{\langle E_2,\sigma \rangle}{}{hang}}
{\begin{matrix}\trans{\langle E_1 + E_2, \sigma \rangle} {}
{hang}\\ \trans{\langle E_2 + E_1, \sigma \rangle} {}
{hang}\end{matrix}} \tag{Nondeterminism$_2$}
\end{gather*}
\caption{Hanging computations in the scc language}
\label{tab:tr-hang}
%\vskip2pt \hrule
\end{acmtable}
\begin{description}
  \item[Nondeterminism] The only case that can lead the system to
  a hanging state is when all the branches are stuck. In this
  case, we can assume no future change of the state will happen
  that give the possibility to the agents to evolve.
\end{description}

To deal with hang states and infinite computations, we enlarged
the observables with the functions
\[
\mathcal{H}_A = \{hang \mid \langle A,\1 \rangle \rightarrow^*
hang\}
\]
that collects all the hang computations of the agent $A$ and
\[
\mathcal{D}_A = \{\bot \mid \langle A,\1 \rangle \text{
diverges}\}
\]
to represent infinite computations.

\section{An Example from the Network Scenario} % Possible Application}
\label{sec:appl}

We consider in this section a simple network
problem, involving a set of processes running on distinct
locations and sharing some variables, %communication channels,
over which they need to synchronize, and we show how to model and solve
such a problem in scc. % their actions.

Each process is connected to a set of variables, shared with other
processes, and it can perform several moves. Each of such moves
involves performing an action over some or all the variables
connected to the process. An action over a variable consists of
giving a certain value to that variable. A special value ``idle''
models the fact that a process does not perfom any action over a
variable. Each process has also the possibility of not moving at
all: in this case, all its variables are given the idle value.

%One of the major concern in this
%framework is to maintain a synchronization between the information
%sent on the shared channels. The aim of the synchronization phase
%between the process is to obtain a state of the network where the

The desired behavior of a network of such processes is that, at
each move of the entire network:
\begin{enumerate}
\item  processes sharing a variable perform the same action over
it;
\item as few processes as possible remain idle.
\end{enumerate}

To describe a network of processes with these features, we use an
SCSP where each variable models a shared variable, and each
constraint models a process and connects the variables
corresponding to the shared variables of that process. The domain
of each variable in this SCSP is the set of all possible actions,
including the idle one. Each way of satisfying a constraint is
therefore a tuple of actions that a process can perform on the
corresponding shared variables.

In this scenario, softness can be introduced both in the domains
and in the constraints. In particular, since we prefer to have as
many moving processes as possible, we can associate a penalty to
both the idle element in the domains, and to tuples containing the
idle action in the constraints. As for the other domain elements
and constraint tuples, we can assign them suitable preference
values to model how much we like that action or that process move.

%penalty level. The idea is to use crisp constraints to model the
%synchronization between the processes, and soft unary constraints
%for the channels, to have as many aliveness
%as possible on the network.

For example, we can use the semiring $S=\langle
[-\infty,0],max,+,-\infty,0\rangle$, where 0 is the best
preference level (or, said dually, the weakest penalty), $-\infty$
is the worst level, and preferences (or penalties) are combined by
summing them. According to this semiring, we can assign value
$-\infty$ to the idle action or move, and suitable other
preference levels to the other values and moves.
%Figure~\ref{fig:processnetwork}(a) shows a possible
%network topology, where each node is a location and each arc
%represent the presence of one or more shared channels between two
%locations.
\begin{figure}[tp]
\centering
%\mbox{\subfigure[A
%network.]{\includegraphics[scale=.6]{processnetwork_a.eps}} \quad
%\subfigure[An SCSP showing the processes and channels located in four
%locations.]{
\includegraphics[scale=.6]{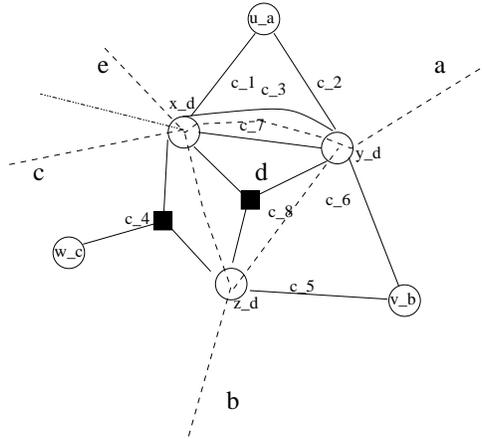} %}}
\caption{The SCSP describing part of a process network}
\label{fig:processnetwork}
\end{figure}
Figure~\ref{fig:processnetwork} gives the details of a part of a
network
% (the one circled in Figure \ref{fig:processnetwork}a)),
and it shows eight processes (that is, $c_1,\ldots,c_8$) sharing a
total of six variables. In this example, we assume that processes
$c_1$, $c_2$ and $c_3$ are located on site {\em a}, processes
$c_5$ and $c_6$ are located on site {\em b}, and $c_4$ is located
on site {\em c}. Processes $c_7$ and $c_8$ are located on site
{\em d}. Site {\em e} connects this part of the network to the
rest. Therefore, for example, variables $x_d$, $y_d$ and $z_d$ are
shared between processes located in distinct locations.

As desired, finding the best solution for the SCSP representing
the current state of the process network means finding a move for
all the processes such that they perform the same action on the
shared variables and there is a minimum number of idle processes.
However, since the problem is inherently distributed, it does not
make sense, and it might not even be possible, to centralize all
the information and give it to a single soft constraint solver.

On the contrary, it may be more reasonable to use several soft
constraint solvers, one for each network location, which will take
care of handling only the constraints present in that location.
Then, the interaction between processes in different locations,
and the necessary agreement to solve the entire problem, will be
modelled via the scc framework, where each agent will represent
the behaviour of the processes in one location.

%The problem can be easily solved as an SCSP by using the
%techniques described in \cite{jacm,padl00,cp2000,BISTATESI}, but
%this needs the presence of an underlying constraint system able to
%solve the problem globally.

%If the processes are not physically on the same location this
%could be difficult or impossible. In fact, the constraint system
%could be able to solve only a local problem, but might not have
%any possibility to know the assignment of the remote channels.

%To solve this problem, we represent each set of processes present
%in a location as an scc agent. E

More precisely, each scc agent (and underlying soft constraint
solver) will be in charge of receiving the necessary information
from the other agents (via suitable asks) and using it to achieve
the synchronization of the processes in its location. For this
protocol to work, that is, for obtaining a global optimal solution
without a centralization of the work, the SCSP describing the
network of processes has to have a tree-like shape, where each
node of the tree contains all the processes in a location, and the
agents have to communicate from the bottom of the tree to its
root. In fact, the proposed protocol uses a sort of Dynamic
Programming technique to distribute the computation between the
locations. In this case the use of a tree shape allows us to work,
at each step of the algorithm, only locally to one of the locations.
In fact, a non tree shape would lead to the construction of non-local
constraints and thus require computations which involve more than one
location at a time.
%
%while the synchronization between processes positioned over different
%locations is guaranteed at the language level by using an agent
%that checks when each of the local computations is finished and then
%posts a
%constraint that synchronize the action performed upon each of the
%remote channels. The computation has to follow a specific order
\begin{figure}[tp]
\centering \mbox{\subfigure[A possible tree structure for our
network.]{\includegraphics[scale=.8]{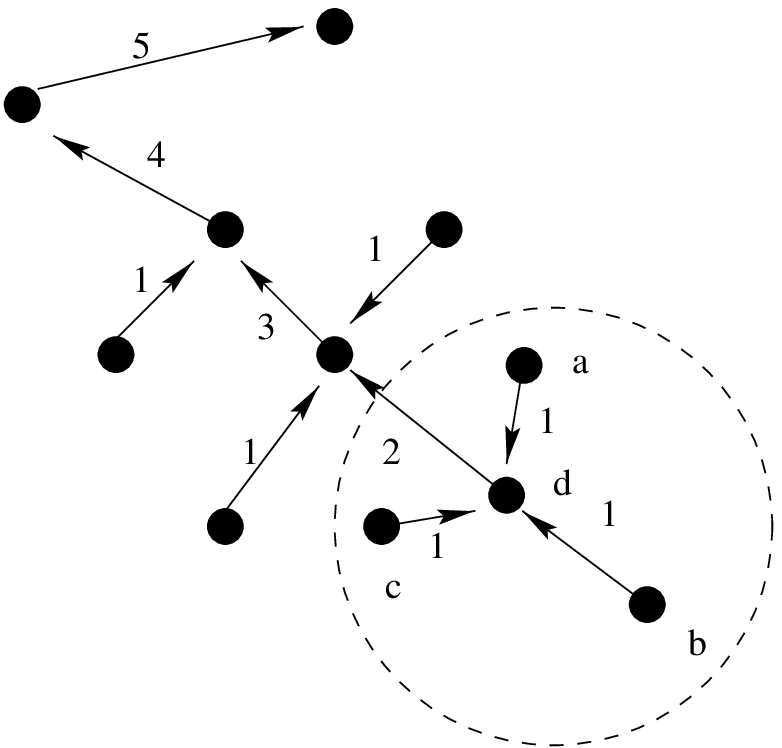}}
\quad \subfigure[The SCSP partitioned over the four
locations.]{\includegraphics[scale=.5]{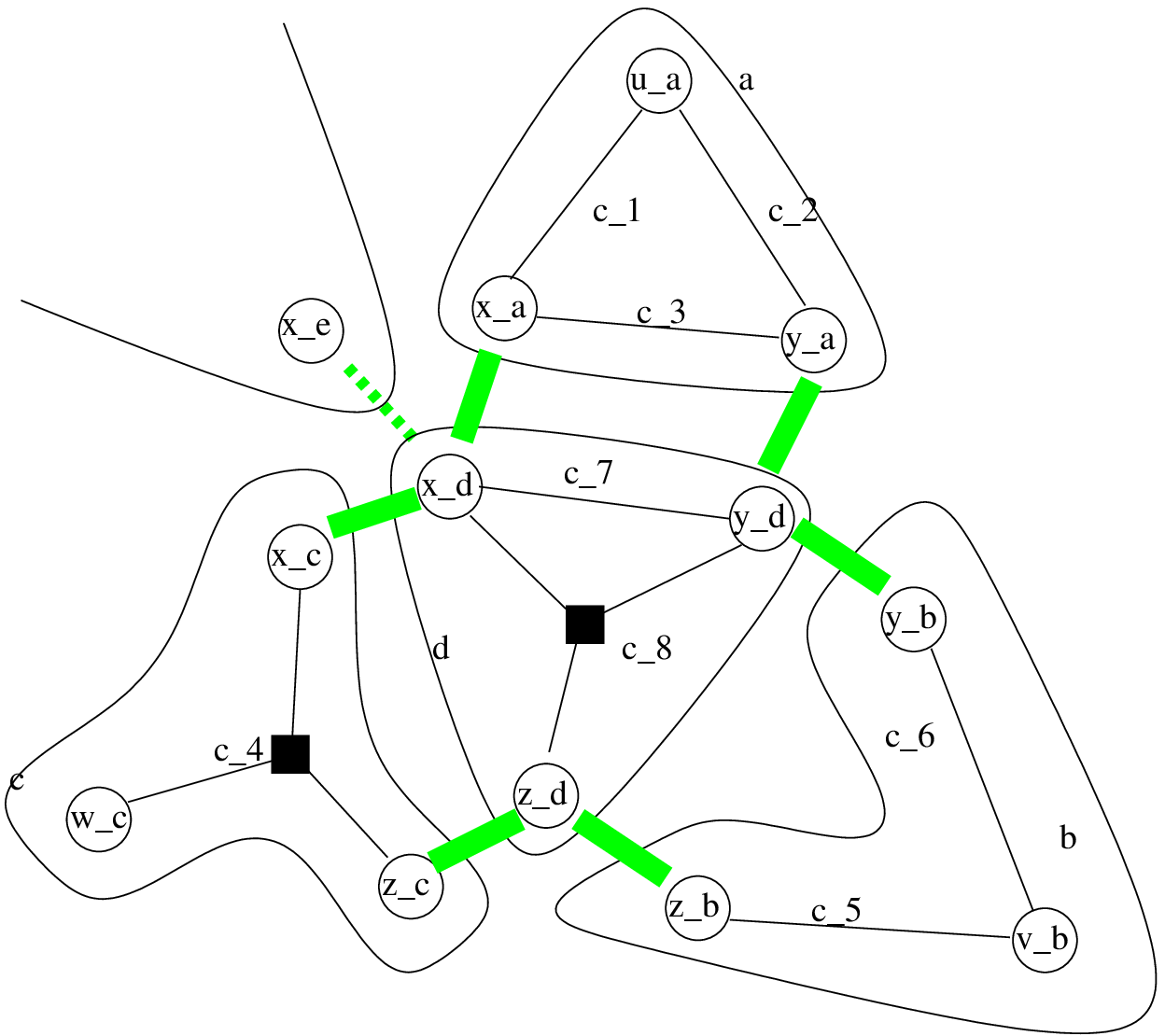}}}
\caption{The ordered process network}
\label{fig:orderedprocessnetwork}
\end{figure}
In our example, the tree structure we will use is the one shown in
%given by a suitable partial order defined on top of the network as
%depicted in
Figure~\ref{fig:orderedprocessnetwork}(a), which also shows the
direction of the child-parent relation links (via arrows).
Figure~\ref{fig:orderedprocessnetwork}(b) describes instead the
partition of the SCSP over the four involved locations.
%to be solved inside each of the locations.
The gray connections represent the synchronization to be assured
between distinct locations. Notice that, w.r.t. Figure
\ref{fig:processnetwork}, we have duplicated the variables
representing variables shared between distinct locations, because
of our desire to first perform a local work and then to
communicate the results to the other locations.

The scc agents (one for each location plus the parallel
composition of all of them)
%running in the locations and the agent in charge
%of final synchronization
are therefore defined as follows:

\begin{equation*}
\begin{split}
A_a& : \exists_{u_a} (tell(c_1(x_a,u_a)\wedge c_2(u_a,y_a)\wedge
c_3(x_a,y_a))\rightarrow tell(end_a=true)\rightarrow stop)
\\
A_b& : \exists_{v_b} (tell(c_5(y_b,v_b)\wedge
c_6(z_b,v_b))\rightarrow tell(end_b=true)\rightarrow stop)
\\
A_c& : \exists_{w_c} (tell(c_4(x_c,w_c,z_c))\rightarrow
tell(end_c=true)\rightarrow stop)
\\
A_d& : ask(end_a=true\wedge end_b=true\wedge end_c=true\wedge
end_d=true)\rightarrow \\
& tell(c_7(x_d,y_d)\wedge c_8(x_d,y_d,z_d) \wedge x_a=x_d=x_c
\wedge y_a=y_d=y_b \wedge z_b=z_d=z_c)
\\
& \rightarrow tell(end_d=true) \rightarrow stop \\
A& : A_a \mid A_b \mid A_c \mid A_d
\end{split}
\end{equation*}

Agents $A_a$,$A_b$,$A_c$ and $A_d$ represent the processes running
respectively in the location $a$, $b$, $c$ and $d$. Note that, at
each ask or tell, the underlying soft constraint solver will only
check (for consistency or entailment) a part of the current set of
constraints: those local to one location. Due to the tree
structure chosen for this example, where agents $A_a$, $A_b$, and
$A_c$ correspond to leaf locations, only agent
%and will not need any check of the variables shared between not
%local processes. The process
$A_d$ shows all the actions of a generic process: first it needs
to collect the results computed separately by the other agents
(via the ask); then it performs its own constraint solving (via a
tell), and finally it can set its end flag, that will be used by a
parent agent (in this case the agent corresponding to location
$e$, which we have not modelled here).

%The technique used in the example can be used in any tree-like
%shape network, in a dynamic programming fashion. The technique,
%really restrict the store to be checked at each tell step.

\section{Conclusions and Future Work}
\label{sec:concl}

We have shown that cc languages can deal with soft constraints.
Moreover, we have extended their syntax to use soft constraints
also to direct and prune the search process at the language level.
We believe that such a new programming paradigm could be very
useful for web and internet programming.

In fact, in several network-related areas, constraints are already
being used
\cite{security-padl01,RFC2702,RFC1102,RajJain00,calisti00}.
%The idea of QoS and of agreement has already been
%extended to the term {\em constraint-based} service and agreement
%\cite{RFC2702,RFC1102,RajJain00,calisti00}.
The soft constraint framework has the advantage over the classical
one of selecting a ``best'' solution also in overconstrained or
underconstrained systems. Moreover, the need to express
preferences and to search for optimal solutions shows that soft
constraints can improve the modelling of web interaction
scenarios.
%
%However, we need to investigate this issue in detail and check the
%usefulness of scc in several web application areas. Thus this
%paper can be regarded as a first step in this direction.

% vedere vari simili lavori:
%\begin{itemize}
%\item dipierro-wickicly
%\item gupta-saraswat
%\end{itemize}

%NOTA: l'avere non funzioni ma valori di semiring potrebbe essere
%una prima estensione al caso classico. Infatti nel caso classico
%non si confrontano sottoinsiemi di tokens per decidere la
%consistenza ma si usa una proprieta' globale dello store (la
%consitenza/inconsistenza). Sembra naturale estendere la nozione di
%consistenza a quella di $\alpha$-consistenza ed avere quindi come
%parametro nella tell ed ask proprio un valore da confrontare con
%una proprieta' globale dello stato che e' l'alpha consistenza.
%Considerare prima caso con valore e poi caso con funzione?

%\section{To do}
%Domanda: Dobbiamo aggiungere assioma sulla commuattivita' e
%associativita del nondeterministic operator $+$? la regola per
%catturare fail e hang e corretta?

\begin{acks}
We are indebted to Paolo Baldan for
valuable suggestions.
\end{acks}

%We are also grateful to the anonymous
%referees for their useful comments on a draft of this paper.

%\begin{spacing}{0.9}
\bibliographystyle{acmtrans}
\bibliography{acronyms,scc}
%\end{spacing}

\begin{received}
...
\end{received}
\end{document}